\documentclass[%
 reprint,
 amsmath,amssymb,
 aps,
prb,
]{revtex4-1}

\usepackage{braket}
\usepackage{graphicx}
\usepackage{dcolumn}
\usepackage{bm}
\usepackage[normalem]{ulem} 
\usepackage{color}

\usepackage{hyperref}
\hypersetup{
    colorlinks,%
    citecolor=blue,%
    linkcolor=blue,%
    urlcolor=blue
}

\newcommand{\bo}{borophene}

\begin{document}

\title{Electronic Stripes and Transport Properties in Borophene 
Heterostructures}

%
\author{G. H. Silvestre and R. H. Miwa}
\affiliation{Instituto de F\'isica, Universidade Federal de Uberl\^andia, \\
        C.P. 593, 38400-902, Uberl\^andia, MG,  Brazil}%
\author{Wanderl\~a L. Scopel}
\affiliation{Departamento de Fı\'{i}sica, Universidade Federal do Espı\'{i}rito 
Santo, 29075-910 Vit\'oria, ES, Brazil.}
\date{\today}

\begin{abstract}

We performed a 
theoretical investigation  of the structural and electronic properties of (i) 
pristine, and  (ii) superlattice structures  of borophene. In (i), by combining 
first-principles calculations, based on the  density functional theory (DFT), 
and simulations of the X-ray Absorption Near-Edge Structure (XANES) we present a 
comprehensive picture  connecting the atomic arrangement  of borophene and the 
X-ray absorption spectra. Once we  have characterized the electronic properties 
of the pristine systems, we next examined the  electronic confinement effects in 
2D borophene superlattices (BSLs) [(ii)]. Here, the BSL structures were made by  
attaching laterally  two different structural phases of borophene. The energetic 
stability, and the electronic properties of those BSLs were examined based on 
total energy DFT calculations. We find a  highly anisotropic electronic 
structure,  characterized by the electronic confinement effects, 
giving rise to ``electronic stripes'', and  
metallic channels ruled by the superlattices. Combining DFT and the 
Landauer-B\"uttiker formalism, we investigate the electronic  transport 
properties in  BSLs. Our results of the transmission probability reveal that the 
electronic transport  is ruled by $\pi$ or a combination of $\pi$ and $\sigma$ 
transmission channels, depending on the atomic arrangement and periodicity of 
the superlattices. Finally, we show that there is huge magnification on the 
directional dependence of the electronic transport properties in BSLs, in 
comparison with the pristine borophene phase. Those findings indicate that BSLs 
 are  quite interesting systems in order to design conductive nanoribbons in a 
2D platform.
\end{abstract}

\maketitle

\section{Introduction}

Since the breakthrough of graphene in 2004,\cite{novoselovScience2004-2} the 
synthesis of two dimensional (2D) materials reached the state of the art. The 
formation of the honeycomb lattice in graphene is dictated   by the C-$sp^2$ 
hybridization of the carbon atoms, favoring the formation of two dimensional  
systems. Such an electronic/orbital  ``ingredient'' is also present in boron 
atoms (B-$sp^2$), and thus pointing out the possibility of synthesis of 2D boron 
systems. 

Indeed, there are several theoretical studies, based on first-principles 
calculations, predicting 2D systems composed by boron 
atoms.\cite{kunstmannPRB2006,tangPRL2007,penevNanoLett2012,wuACSNano2012,
xuNanoscale2018} Meanwhile, in a seminal work, Mannix {\it et 
al.}\cite{mannixScience2015} obtained 2D boron sheets (borophene) on the Ag(111) 
surface;  and few months latter, Feng {\it et al.}\cite{fengNatChem2016} 
synthesized two new different structural phases of  borophene on the same 
substrate. The former boron 2D structure is characterized by a buckled geometry  
where  the boron atoms are six-fold coordinated; while  the latter presents a 
planar geometry with the presence  boron vacancies, giving rise to  six-, five-, 
and four-fold boron coordinations. In a very recent work, Rajan {\it et 
al.}\cite{ranjanAdvMat2019}  successfully performed a large scale synthesis of
free-standing borophene sheets, which is a quite important step towards the 
design layered devices based on 
borophene.\cite{zhangNanoscale2016,jiangNanoEnergy2016}

The combination of materials with distinct electronic properties has been one of 
the most promising route  to the design new electronic devices. Such a material 
engineering began with 3D semiconductor heterostructures, like (GaAs)$_{\rm 
m}$/AlAs$_{\rm n}$ superlattices 
(SLs).\cite{esaki1970superlattice,bylanderPRB1987,hsuPRB1994} The electronic 
properties of those hybrid systems are mostly ruled by the  intrinsic 
characteristics of the pristine materials (e.g. GaAs and AlAs), and their 
respective size/periodicity (m and n) within the SL.  Upon further progress on 
the growth processes, 2D lateral heterostructures have been successfully 
synthesized, like  
WS$_2$/MoS$_2$,\cite{gongNatMat2014,huangNatMat2014,zhangNanoLett2014, 
yooJACS2015,chenACSNano2015} and 
2H-MoS$_2$/1T-MoS$_2$\cite{edaNanoLett2011,edaACSNano2012}. The former is a 2D 
heterostructure composed by different materials (WS$_2$ and MoS$_2$), while in 
the latter we have a single material but with  different structural phases (2H 
and 1T).

Taking advantage of the  polymorphism of 2D  boron 
lattices,\cite{penevNanoLett2012,wuACSNano2012,xuNanoscale2018} Liu {\it et 
al.}\cite{liuNatMat2018} built-up borophene heterostructures composed by 
self-assembled periodic structures of linear rows with  different atomic 
arrangements. In particular, borophene phases with hollow hexagons concentration 
($\eta$) of 1/6 and 1/5, so called S1 and S2.  As depicted in 
Fig.\,\ref{models0},  each borophene phase is characterized by  different 
electronic structures and density of states near the Fermi level, which can be 
exploited in order to design new electronic devices in  2D platforms. For 
instance,  promoting the electronic confinement, and strengthening the 
directional dependence of the electronic transport and 
properties.\cite{padilhaPCCP2016,pengJMatChem2016,zengJPhysChemC2019} 

In this work, based on the {\it first-principles} density 
functional theory (DFT) calculations,  we performed a theoretical study of (i) 
pristine borophene sheets, and (ii) borophene heterostructures composed by 
combinations of S0, S1 and S2 phases, {\it viz.}: S0$_{\rm m}$/S1$_{\rm n}$, 
S0$_{\rm m}$/S2$_{\rm n}$, and  S1$_{\rm m}$/S2$_{\rm n}$ forming  borophene 
superlattices (BSLs) with m/n periodicities. In (i), combining   DFT results and 
simulations of the X-ray absorption spectra (XAS), we characterize the 
electronic--structural properties of the pristine 
phases.  In (ii) we examined the electronic confinement in BSLs, where 
we show the formation of electronic stripes ruled by borophene rows.
Moreover, combining DFT calculations and  the Landauer-B\"uttiker 
formalism,\cite{buttikerPRB1985,meirPRL1985} we studied the electronic 
transport properties in the BSLs as a function of the m/n periodicity.

\section{Method}

The calculations were performed within the density  functional theory, where the 
exchange-correlation term was described within the generalized gradient 
approximation as proposed by   Perdew, Burke and Ernzerhof (GGA-PBE).\cite{PBE} 
The calculations of the  equilibrium geometries, total energies, and the 
electronic band structures were done by using  the VASP code.\cite{vasp1,vasp2}. 
In order to satisfy the periodic  boundary conditions,  we have 
used the super-cell approach with a vacuum region of 15\,\AA\ perpendicularly to 
the borophene sheet. Kohn-Sham (KS) orbitals were expanded in a plane wave 
basis set with an energy cutoff of 400 eV. The 2D Brillouin Zone (BZ) is sampled 
according to the Monkhorst-Pack method,\cite{mp} using a gamma-centered 
12$\times$12$\times$1 mesh. The electron-ion interactions are taken into account 
using the Projector Augmented Wave (PAW) method.\cite{paw}  
Atomic positions and cell-volume were fully relaxed, 
considering a convergence for the atomic forces on each atom smaller than 
$0.025$\,eV/{\AA}. Those relaxed geometries were used for  the XAS and 
electronic transport calculations described below.

The boron K-edge X-ray absorption near-edge structure (XANES) spectra were
simulated using the theoretical approach implemented in the XSpectra 
code,\cite{xas1,xas2,xas3} supplied with QUANTUM ESPRESSO.\cite{espresso}
We have used ultrasoft pseudopotentials, where  in order to describe the K-edge 
spectra, we built a pseudopotential with a core hole in the 1$s$ orbital, and 
the all electron wave function were recovered by using the  GIPAW\cite{gipaw} 
approach. We have considered a set of 12$\times$12$\times$1 
k-points to the BZ sampling,   energy cutoffs of  40\,Ry for the plane wave 
basis set (to expand the KS orbitals) and 400\,Ry for the self-consistent total 
charge density.

The electronic transport calculations was explored using the non-equilibrium 
Green's function (NEGF) formalism using the DFT Hamiltonian as implemented in 
the TranSiesta\cite{siesta,transiesta} code. The KS orbitals 
were expanded in  a linear combination of numerical pseudo-atomic orbitals using 
 split-valence double-zeta basis set including polarization 
functions.\cite{dzp, EShiftSiesta}  The BZ samplings were performed using two 
different set of k-point meshes, 200$\times$12$\times$1 or 
12$\times$200$\times$1 according with the  electronic transport  directions.

The total  transmission probability of electrons with energy $E$ [$T(E)$] from 
the left electrode to reach the right electrode passing through the scattering 
region is given by,

\begin{equation}
T\left(E \right) = Tr \left[\Gamma_{\mathrm R}\left(E,V\right) G^{\mathrm 
R}\left(E,V\right) \Gamma_{\mathrm L}\left(E,V\right) G^{\mathrm 
A}\left(E,V\right) \right] ~,
\end{equation}
where $\Gamma_{L,(R)} \left(E,V\right)$ is the coupling with 
the left and right electrodes and $G^{R,(A)}$ is the retarded 
(advanced) Green function matrix of the scattering region. 
The current I is evaluated by using Landauer-B\"uttiker 
formula,\cite{buttikerPRB1985,landauer1988generalizd}

\begin{equation}
I\left(V\right)=\frac{2e}{h}\int 
T\left(E,V\right)\left[f\left(E-\mu_L\right)-f\left(E-\mu_R\right) 
\right]dE,
\end{equation}
where $f\left(\epsilon\right)$ is the Fermi-Dirac distribution for energy 
$\epsilon$ and $\mu_{L(R)}$ is the electrochemical potential of left 
(right) electrodes.

\section{Results and Discussions}

\begin{figure}[h]
\includegraphics[width=\columnwidth]{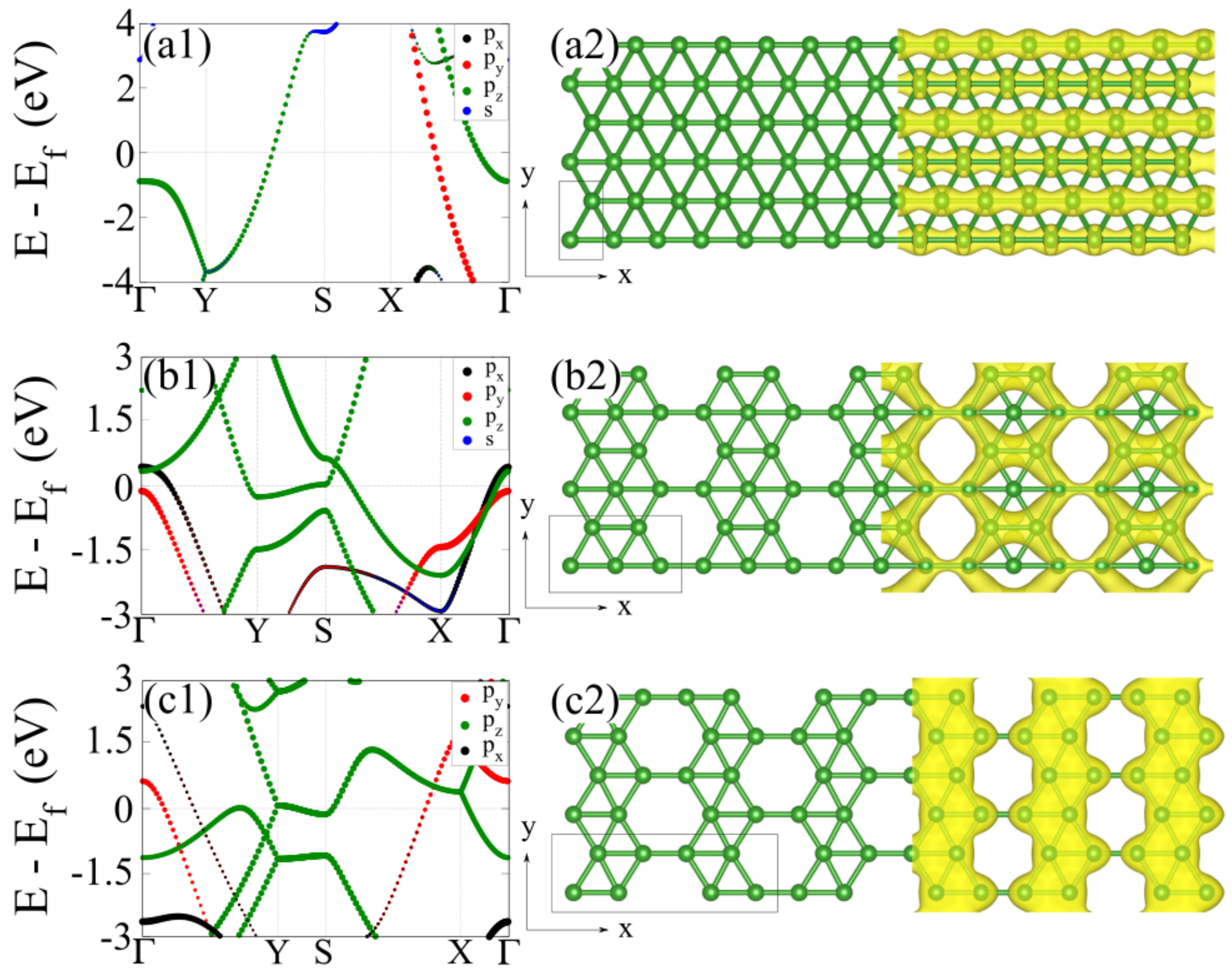}
\caption{\label{models0} Electronic structure, and structural models of the 
pristine borophene sheets with different vacancy concentration ($\eta$), (a) S0 
($\eta=0$), (b) S1 ($\eta=1/6$), and (c) S2 ($\eta=1/5$); and the projected 
density of states near to Fermi level, $E_{\rm F}\pm 0.1$\,eV. The 
isosurfaces 
in (a2) and (b2) are equal to $0.0025\,e/$\AA$^3$, and in (c2) 
$0.005\,e/$\AA$^3$.}
\end{figure}

\subsection{Pristine sheets}

In Fig.\,\ref{models0} we present the structural models and the electronic bands 
structures  of pristine \bo\, S0 [Fig.\,\ref{models0}(a)], S1 
[Fig.\,\ref{models0}(b)], and S2 [Fig.\,\ref{models0}(c)] phases. In S0 the B 
atoms are six-fold coordinated, and present a buckled geometry giving rise to 
two layers of boron atoms with a vertical distance ($z$ direction) of 0.87\,\AA. 
The electronic band structure of S0 is highly anisotropic, characterized by the 
formation of metallic bands for wave vectors along the $\Gamma$-X and Y-S 
directions [Fig.\,\ref{models0}(a1)].\cite{mannixScience2015} In 
Fig.\,\ref{models0}(a2)  we present  projected the electronic density of states 
near the Fermi level, $E_{\rm F}\,\pm\,0.1$\,eV, where we can identify the wave 
function overlap along the $x$-direction. Indeed, in a recent work, we verified 
that the electronic transport properties in S0 are characterized by a strong 
directional dependence.\cite{padilhaPCCP2016} Upon the presence of vacancies in 
S0, as observed in S1 and S2, the vertical buckling  has been 
suppressed.\cite{fengNatChem2016} The planar boron sheets, with four- five- and 
six-fold coordinated B atoms,  are characterized by a density of hollow hexagons 
($\eta$)\,\cite{tangPRL2007} of  1/6 and 1/5, respectively.\cite{hole} The 
energetic stability of those pristine systems was examined through the 
calculation of the cohesive energy ($E^c$).\cite{CohesionEnergy} We found $E^c$ 
of 6.21, 6.25, and 6.27\,eV/atom for S0, S1 and S2 phases. The energetic 
preference of S1 and S2, compared with S0 is in agreement  with the previous 
{\it first-principles} 
results.\cite{wuACSNano2012,mannixScience2015,fengNatChem2016} 

\begin{figure}[h]
\includegraphics[width=\columnwidth]{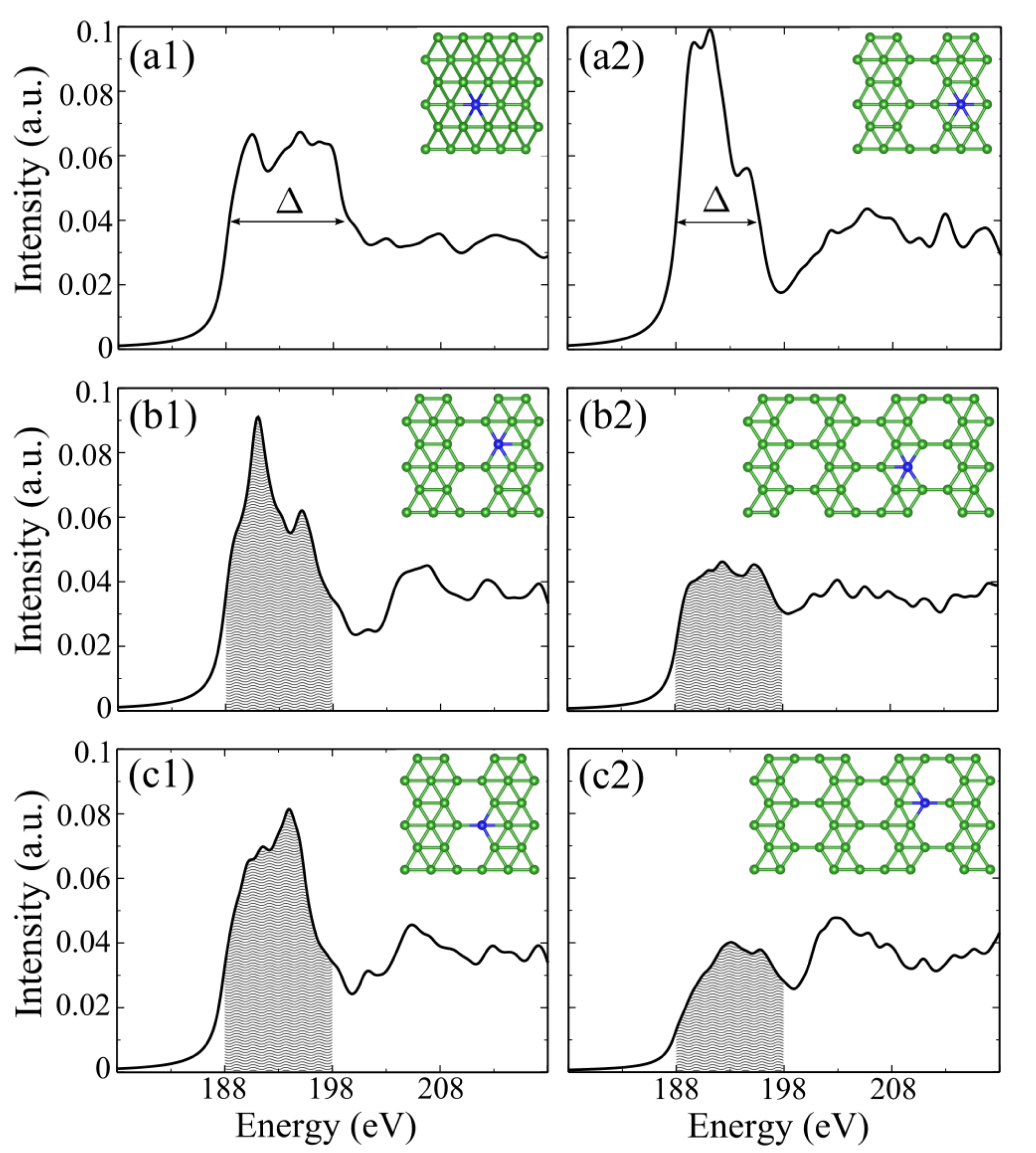}
\caption{\label{xas} The simulated X-ray Absorption Near-Edge Structure (XANES) 
spectra of the pristine borophene sheets, K-edge spectra of six-fold boron 
atoms  
in S0 (a1) and  S1 (a2); five-fold coordinated boron  atoms in S1 (b1) and S2 
(b2); four-fold coordinated boron  atoms in S1 (c1) and S2 
(c2). Inset, the local geometry of the probed boron atoms (blue circles).}
\end{figure}

The metallic character  of borophene in S1 and S2 phases  has been kept, 
Figs.\,\ref{models0}(b1) and (c1), however, compared with S0, the electronic 
anisotropy is somewhat dimmed. We find that the metallic bands of S1 are mostly 
composed by the B-2$p_z$ and -2$p_x$ orbitals localized on the four- and 
five-fold coordinated B atoms. Meanwhile, in S2 the formation of the metallic 
bands along the $\Gamma$-Y and  S-X directions is dictated by a combination  of 
$\sigma$- (B-$2p_y$) and $\pi$-orbitals (B-$2p_z$).

In order to get some insight into the structural and electronic features of the 
borophene sheets,  we simulate the boron K-edge X-ray absorption 
near-edge structure (XANES).  By considering the polarization vector 
perpendicular to the borophene sheet ($\hat\varepsilon_\perp$),  we have 
examined  the main features associated with the 
B-$1s$\,$\rightarrow$\,$\pi^\ast$ transitions.  

Boron atoms in S0, S1 and S2 are characterized by different bonding geometries 
and coordinations. (i) In S0 and S1 we have six-fold coordinated boron atoms. 
The former presents a buckled geometry, giving rise to boron lines along the 
$x$-direction [Fig.\,\ref{models0}(a2)], while S1 exhibits a planar geometry, 
where the six-fold boron atoms are separated by boron vacancy lines, 
Fig.\,\ref{models0}(b2). As shown  in Fig.\,\ref{xas}(a1) and (a2), those 
structural differences can be identified at the near-edge absorption interval; 
namely, the B-$1s$\,$\rightarrow$\,$\pi^\ast$ absorption peak in S0  present  a 
larger energy dispersion [$\Delta$ in Fig.\,\ref{xas}(a1)] in comparison with 
its counterpart in S1 [Fig.\,\ref{xas}(a2)]. Thus, indicating that the presence 
of vacancy lines, and the planar geometry of S1 result in a more localized 
$\pi^\ast$ states in comparison with the buckled S0 phase. At 
the atomic scale,  we have a nice manifestation of the role played by the local 
geometry on the absorption spectra. In S0 the B--B bond length ($d$)  between B 
atoms lying on the same layer, $d$ = 1.61\,\AA, is shorter than the ones of the  
six-fold coordinated  boron atoms in S1 ($d$ = 1.71\,\AA), and thus  increasing 
the  $\pi^\ast$  hybridizations in S0 in comparison with S1. (ii) S1 and S2 
are planar structures, where we have vacancies, five-fold, and four-fold 
coordinated boron atoms. The former is characterized by  hexagonal boron 
structures separated by vacancy lines, while in S2  we have zigzag rows of 
boron-dimers ($d$ = 1.67\,\AA) separated by vacancy lines. As 
shown in Fig.\,\ref{models0}(b2) and (c2), the electronic density of states of 
S1 and S2, near the Fermi level, present  different features. The hybridization 
in S1 somewhat mimic the one present in graphene, while in S2  there is charge 
density overlap along the zigzag rows of B-dimers.  Our simulated XANES results 
capture those structural and electronic differences. In S1, the near edge 
spectra [shaded regions in Fig.\,\ref{xas}(b1) and (c1)] are characterized by a 
more intense and less dispersive  B-$1s$\,$\rightarrow$\,$\pi^\ast$ transitions 
when compared with the ones in S2, [shaded regions in Figs.\,\ref{xas}(b2) and 
(c2)]. Thus, indicating that the presence hexagonal (dimer) 
structures in S1 (S2) gives rise to a more (less) localized character of the 
$\pi^\ast$ orbitals near the Fermi level in S1 (S2).

Summarizing,  firstly we have charaterized the structural and 
electronic properties of pristine S0, S1, and S2 borophene phases, and further 
XANES simulations revealed the connection between the local atomic structure, 
{\it i.e.} coordination and equilibrium geometry, and the X-ray absorption 
features.
Those findings provide not only  fingerprints for the different borophene 
structural phases, but also key insights into experimental measurements.
 
\begin{figure}[h]
\includegraphics[width=\columnwidth]{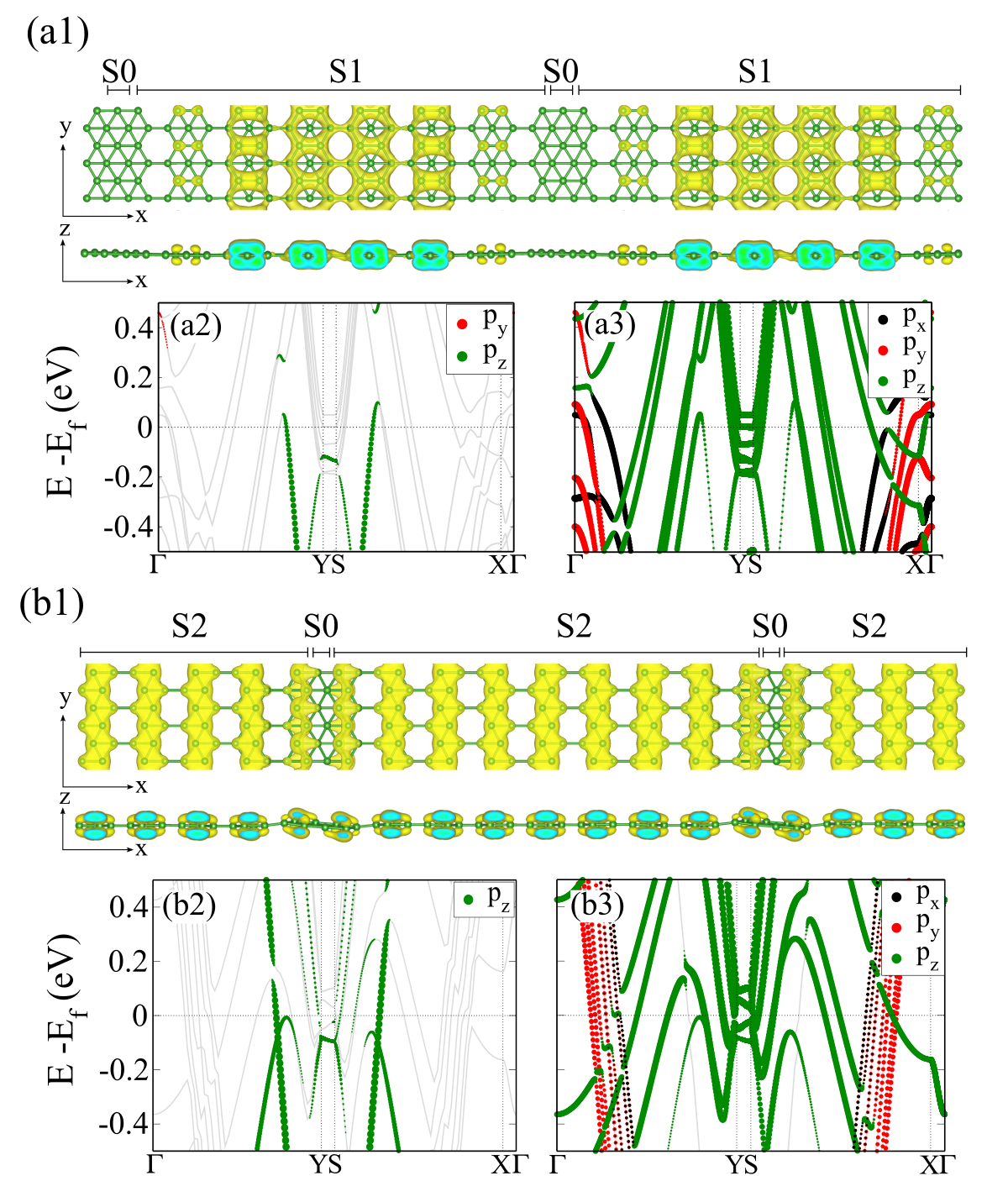}
\caption{\label{bandas-proj-mix} Structural models (top-view  and side-view) and 
the projected electronic density of states near the Fermi level, $E_{\rm F}\pm 
0.1$\,eV, of S0$_1$/S1$_7$ (a1), and S0$_2$/S2$_4$ (b1) BSLs. Electronic band 
structure of  S0$_1$/S1$_7$ [S0$_2$/S2$_4$] projected on  S0 (a2) [(b2)] and S1 
(a3) [S2 (b3)] regions. Isosurfaces of  0.003\,$e$/\AA$^3$.}
\end{figure}

\begin{figure*}[h]
\centering
\includegraphics[width=2\columnwidth]{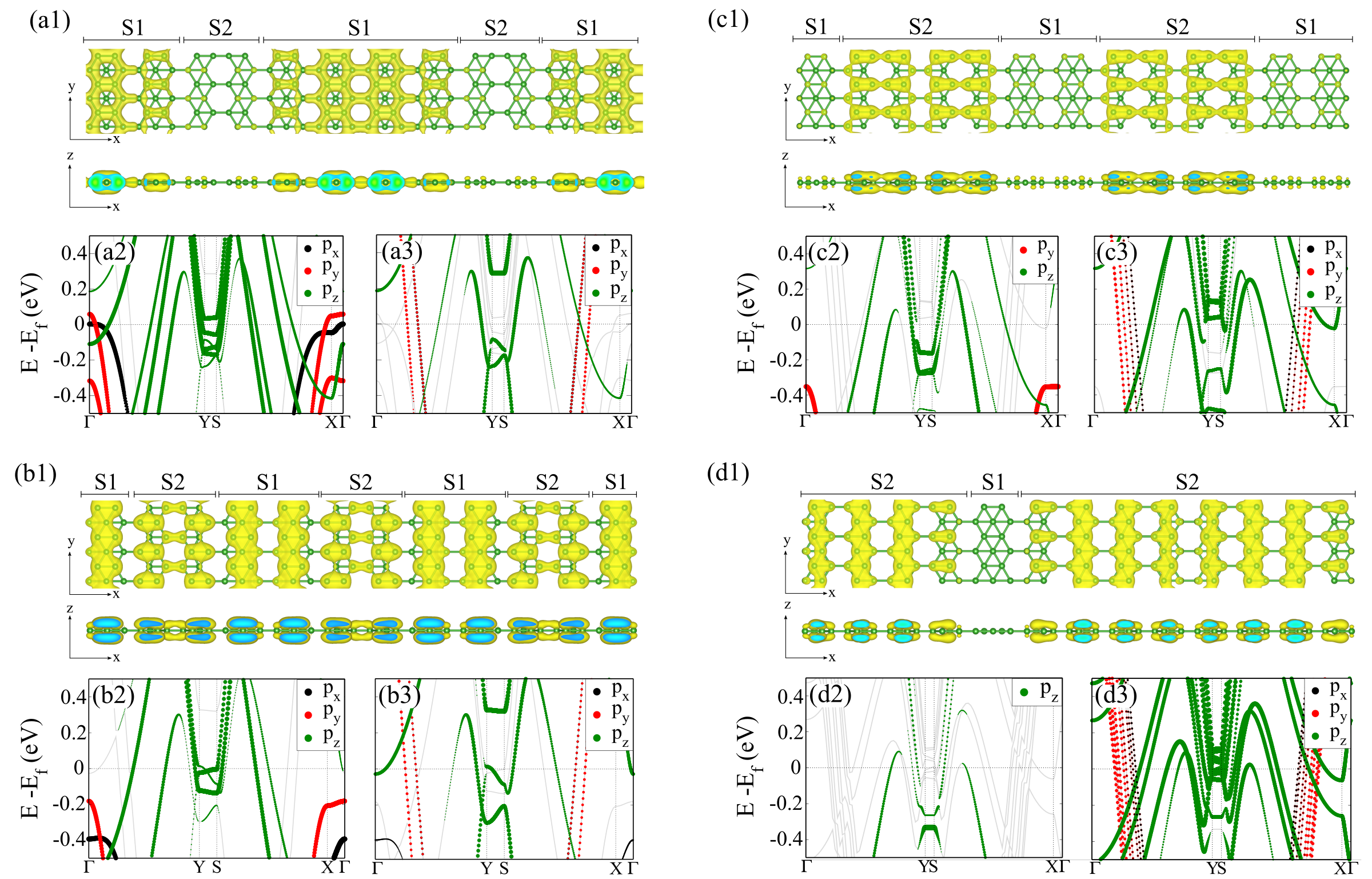}
\caption{\label{bandas-proj-s1s2-c} Structural models (top-view and side-view)  
and the electronic density of states within $E_{\rm F}\pm0.1$\,eV, of  
S1$_4$/S2$_1$ (a1), S1$_2$/S2$_1$ (b1), S1$_2$/S2$_2$ (c1), and  S1$_1$/S2$_4$ 
(d1) BSLs. Electronic structure and the projection of the energy  bands on  
S1$_4$ (a2) and S2$_1$ (a3); S1$_2$ (b2) and S2$_1$ (b3); S1$_2$ (c2) and S2$_2$ 
(c3);  S1$_1$ (c2) and S2$_4$ (c3). Isosurfaces of  0.003\,$e$/\AA$^3$ in (a) 
and (d); and 0.0015\,$e$/\AA$^3$ in (b) and (c).}
\end{figure*}

\subsection{2D boron superlattices}

Motivated by the recent advances on the synthesis of 2D 
heterostructures,\cite{liuNatMat2018} and the theoretically proposed electronic 
localization upon the presence of defects in 
borophene,\cite{kistanovNanoscale2018} we examined the structural and electronic 
properties of borophene lateral heterostructures. In particular, we have 
considered 2D borophene superlattices (BSLs) combining different structural 
phases of borophene, with m/n periodicities,  separated by zigzag edges. For 
instance, in S0$_1$/S1$_7$ (m=1/n=7) we have one periodic unit cell of S0 
separated by seven periodic unit cells of S1, Fig.\,\ref{bandas-proj-mix}(a1). 
It is worth noting that for such a lower proportion of S0, the 
vertical buckling in S0 has been suppressed. In S0$_1$/S1$_7$ the B--B bond 
length between the B atoms lying on the same (different) layer(s) increases 
(reduces) from 1.61\,\AA\ (1.87\,\AA) in pristine S0, to 1.83\,\AA\ (1.73\,\AA) 
at the inner sites of the S0 region. Similarly in S0$_1$/S2$_4$, 
Fig.\,\ref{bandas-proj-mix}(b1), where the vertical buckling at the inner sites 
of  S0 reduces to 0.03\,\AA, with  bond length between the B atoms lying on the 
same (different) layers(s) of 1.72\,\AA\ (1.70\,\AA). The energetic stability 
of those BSLs was inferred by the calculation of the cohesive energies ($E^c$), 
where we found $E^c$ of 6.26\,eV/atom. Our results of $E^c$,  summarized in 
Table~I, show that the cohesive energies of the other BSLs are  practically the 
same when compared with the ones of the pristine phases. Thus, supporting the 
energetic stability of those boron superlattices.

\begin{table}[h]
\centering
\caption{\label{binding} Cohesive energies ($E^c$) of the BSLs (eV/atom) with 
different m/n periodicities, namely S0$_{\rm m}$/S1$_{\rm n}$, S0$_{\rm 
m}$/S2$_{\rm n}$, and S1$_{\rm m}$/S2$_{\rm n}$.}
\begin{tabular}{lllll}
\hline
\hline
 \multicolumn{5}{c}{\underline{S0$_{\rm m}$/S1$_{\rm n}$}} \\
 m/n   & 1/7    & 2/6    &    3/5    &  4/4  \\   
 $E^c$ & 6.26   & 6.25   &    6.25   &   6.23 \\
 \hline
 \multicolumn{5}{c}{\underline{S0$_{\rm m}$/S2$_{\rm n}$}} \\
 m/n   & 2/4   & 2/2    &  4/1      &  8/1    \\
 $E^c$ & 6.27   &  6.27  &  6.23     &  6.23  \\
 \hline
  \multicolumn{5}{c}{\underline{S1$_{\rm m}$/S2$_{\rm n}$}} \\
 m/n   & 4/1    & 2/1    &  2/2  &  1/4   \\
 $E^c$ & 6.26   &  6.27  & 6.27      &  6.27   \\
 m/n   & 6/1    &  1/1   &  1/2      &  1/6    \\
 $E^c$ & 6.26   &  6.27  &  6.27     &  6.27   \\
 \hline
 \hline
\end{tabular}
\end{table}

\begin{figure}[ht]
\includegraphics[width=\columnwidth]{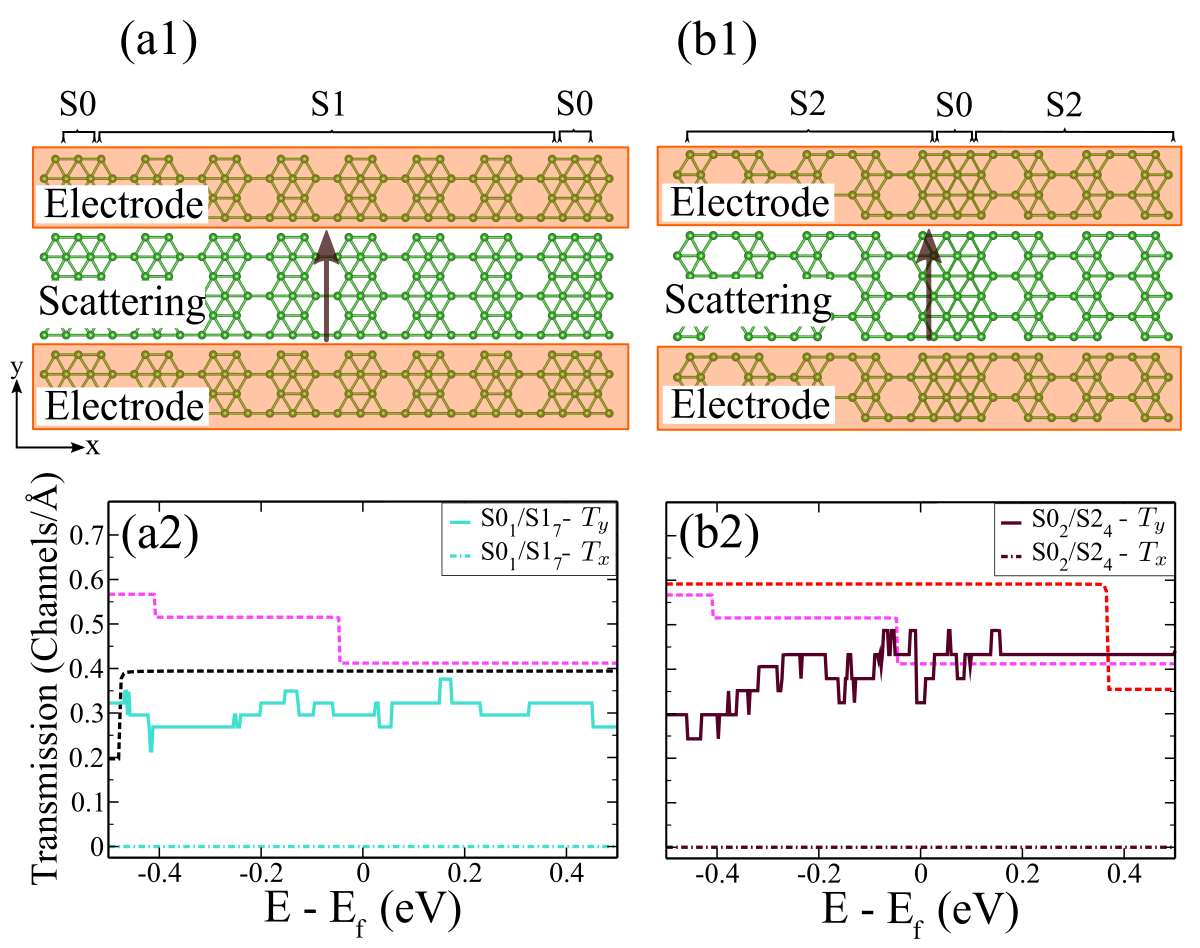}
\caption{\label{trans-s0} Setup used for the electronic transport calculations 
 along the $y$-direction of S0$_1$/S1$_7$ (a1), 
and S0$_2$/S2$_4$ (b1).  Transmission  coefficients [T(E)] of S0$_1$/S1$_7$ 
(a2) and S0$_2$/S2$_4$ (b2). Pink, black and red dashed lines indicate the 
transmission coefficients of S0, S1, and S2 pristine phases.}
\end{figure}

Next we examine the electronic properties of the BSLs as a function of the 
relative width (m/n) of each  structural phase. In S0$_1$/S1$_7$ the 
electronic states near the Fermi level, $E_{\rm F}\pm 0.1$\,eV, are mainly 
localized on S1  [Fig.\,\ref{bandas-proj-mix}(a1)], giving rise 
to electronic stripes composed by graphene-like  $\pi$-hybridizations along the 
four- and five-fold coordinated boron atoms, of the inner sites of S1, 
separated by S0$_1$ rows. The localization of the electronic states can be 
also identified through the projection of the  energy bands on S0 and S1,  
Fig.\,\ref{bandas-proj-mix}(a2) and (a3), respectively; where  it is noticeable 
the most of metallic bands lying on S1. Such a localization of the electronic 
states  near the Fermi level, and the electronic separation of the metallic 
bands have been also observed in  S0$_2$/S2$_4$, Fig.\,\ref{bandas-proj-mix}(b). 
Here, the  electronic states are predominantly localized along the  S2 region. 
However, as shown in Fig.\,\ref{bandas-proj-mix}(b3), in addition to the $\pi$ 
orbitals, we find that the $\sigma$-hybridizations also contribute to the 
formation of the metallic bands. The localization of the 
electronic states in the other S0$_{\rm m}$/S1$_{\rm n}$ and S0$_{\rm 
m}$/S2$_{\rm n}$ combinations,  indicated in Table\,\ref{binding},  are 
presented  in the Appendix, Figs.\,\ref{band-proj-2-1} and \ref{band-proj-3}.

The large number of possible structural combinations to build up 
heterostructures, based on borophene,  is a  interesting  degree of freedom in 
order to control/design the electronic properties  in 2D systems. Very recently, 
borophene heterostructures composed  by  S1 and S2  phases have been 
successfully synthesized,\cite{liuNatMat2018} where the authors identified S1/S2 
periodic assemblies with different relative concentrations.  Here we will 
examine the S1$_{\rm m}$/S2$_{\rm n}$ BSLs, with m/n of 4/1, 2/1, 2/2, and 1/4, 
depicted in Fig.\,\ref{bandas-proj-s1s2-c}. At the equilibrium geometry, (i) the 
planar structure has been preserved, and (ii) the S1/S2 zigzag interface boron 
atoms are neatly arranged, where the B--B bond lengths and angles  are 
practically the same compared with the ones  the pristine structures. 
Indeed, at the equilibrium geometry, the bond lengths and 
angles at the interface region changes by less than 0.8\% and 1.5\%, 
respectively, compared with the ones of the pristine S1 and S2 phases.   Thus, 
indicating that S1/S2 interfaces present lower strain than in  
S0/S1 and S0/S2.

\begin{figure}[ht]
\includegraphics[width=\columnwidth]{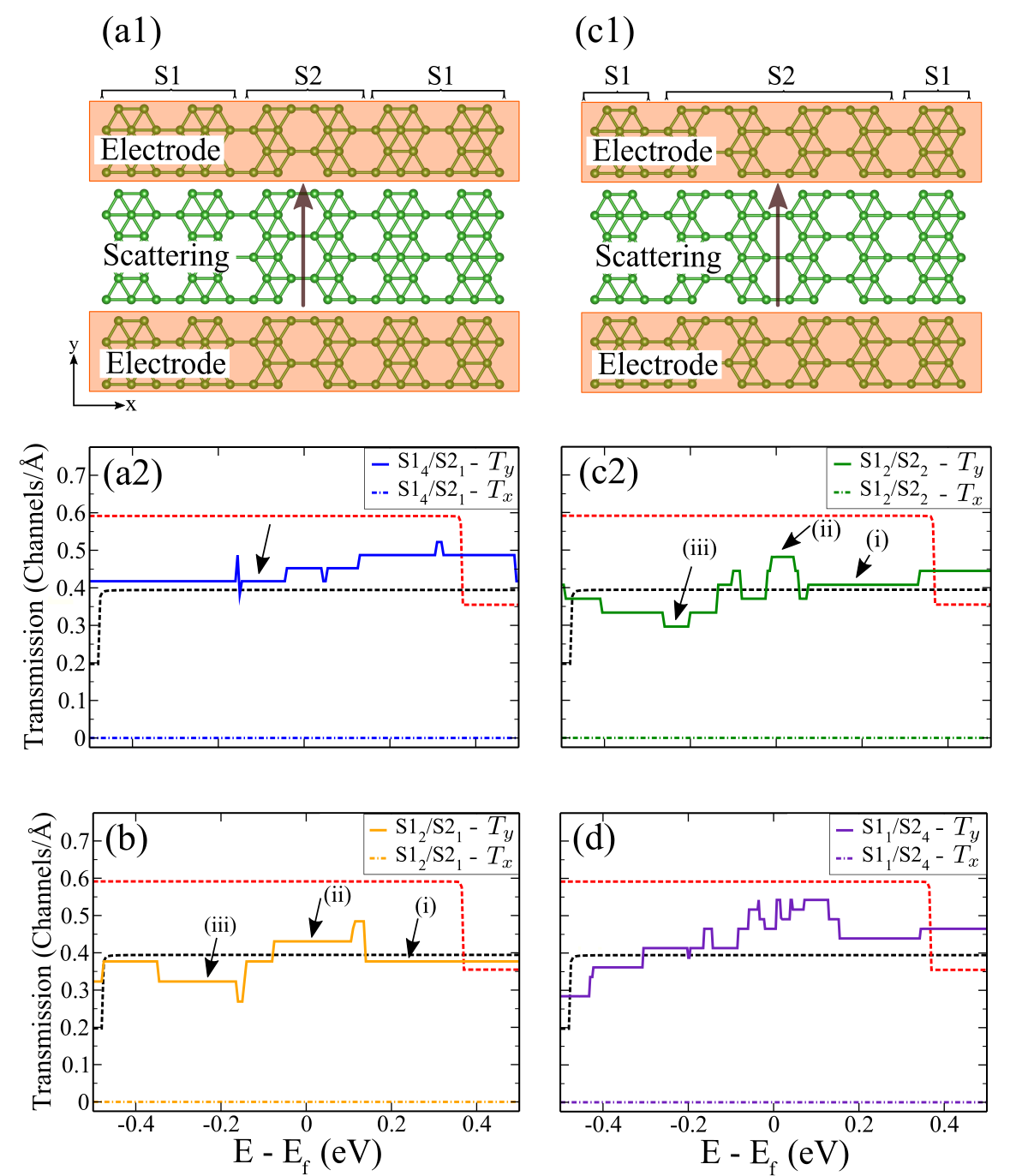}
\caption{\label{trans-s1s2}Setup used for the electronic transport calculations 
along the $y$-direction of S1$_4$/S2$_1$ (a1), and S1$_2$/S2$_2$  (c1). 
Transmission probabilities [T(E)] of S1/S2 BSLs, S1$_4$/S2$_1$ (a2),
S1$_2$/S2$_1$ (b), S1$_2$/S2$_2$ (c2), S1$_1$/S2$_4$ (d). Black and red dashed 
lines indicate the transmission coefficients of the pristine S1 and S2 phases.}
\end{figure}

In S1$_4$/S2$_1$, the electronic states within $E_{\rm F}\pm 0.1$\,eV are mostly 
confined in S1 ruled by  S2, 
\textcolor{blue}{Fig.\,\ref{bandas-proj-s1s2-c}(a1)}. There is a charge density 
overlap along  S1, giving rise to metallic bands for wave vectors parallel to 
the $\Gamma$Y and SX directions, Fig.\,\ref{bandas-proj-s1s2-c}(a2) and (a3). 
Those metallic bands are predominantly composed by  B-$2p_z$ orbitals lying 
on the  four- and five-fold coordinated boron atoms of S1.  Meanwhile, the 
projection of the energy bands on S2 reveals contributions from B-$2p_z$ and 
-$2p_y$ orbitals to the formation of the metallic bands. On the other hand, the 
electronic states with wave vector perpendicular to the S1/S2 interface, namely  
$\Gamma$X and YS directions, are characterized by dispersionless energy bands  
localized mainly on S1.  The localization of the electronic 
states on the other S1$_{\rm m}$/S2$_{\rm n}$ BSLs,  m/n = 6/1, 1/1, 1/2, and 
1/6 (Table\,\ref{binding})   are presented in the Appendix 
[Fig.\,\ref{banda-proj-4}], where we confirm the formation of 
tuneable electronic stripes in S1/S2 BSLs.

Reducing the width of S1, S1$_4$/S2$_1$\,$\rightarrow$\,S1$_2$/S2$_1$, the 
distribution of the  electronic density of states near the Fermi level, on S1 
and S2, becomes almost equivalent, however they present different features, 
Fig.\,\ref{bandas-proj-s1s2-c}(b1). The metallic bands   are characterized by 
$\pi$-hybridization  through S1, while in S2 we have both, {\it i.e.} $\pi$- and 
$\sigma$-hybridizations (B-$2p_y$),  with a major contribution from the latter 
along the  zigzag B-dimers lines. Keeping the width of S1, and increasing S2, 
S1$_2$/S2$_1$\,$\rightarrow$\,S1$_2$/S2$_2$, we find that the electronic states 
near the Fermi level  become more localized on S2, mostly on  the four-fold 
coordinated boron atoms, as shown in Fig\,\ref{bandas-proj-s1s2-c}(c1). Indeed, 
the $\pi$-and $\sigma$-hybridizations along the S2 rows have been strengthened,  
giving rise to metallic bands along the $\Gamma$Y and SX directions, 
concomitantly there is a reduction of the electronic contribution from S1. 
Finally, upon further reduction of S1 and increase of S2, 
S1$_2$/S2$_2$\,$\rightarrow$\,S1$_1$/S2$_4$ [Fig.\,\ref{bandas-proj-s1s2-c}(d)], 
there is  a noticeable change on the electronic distribution along  S2, where 
the wave function overlap between the B-$2p_y$ orbitals becomes more intense,  
strengthening the   $\sigma$-hybridizations and the orbital localization along 
S2,  characterized by zigzag rows of B-dimers. It is worth 
noting that   the  B-dimer bond length in S2$_4$ ($d$ = 1.67\,\AA) is 
the same compared with that of S2 pristine.

Focusing on the electronic transport properties in 2D systems, its 
``directional dependence'', and the ``role of intrinsic line defects'' are 
important issues that have been addressed in recent 
studies.\cite{padilhaPCCP2016,shuklaPCCP2018,zengJPhysChemC2019, 
sivaramanJPhysChemC2016} Here, we  discuss the role  of the 
electronic confinement on the  transport properties along the BSLs S0/S1, S0/S2 
[Fig.\,\ref{bandas-proj-mix}] and S1/S2 [Fig.\,\ref{bandas-proj-s1s2-c}]. We 
have examined the electronic transport  along the superlattices, {\it i.e.} 
along the $y$-direction. Perpendicularly to the interface, 
$x$-direction, we obtained nearly zero transmission probabilities, 
thus indicating that the electronic transport in BSLS presents a strong 
directional dependence. 

\begin{figure}[ht]
\includegraphics[width=\columnwidth]{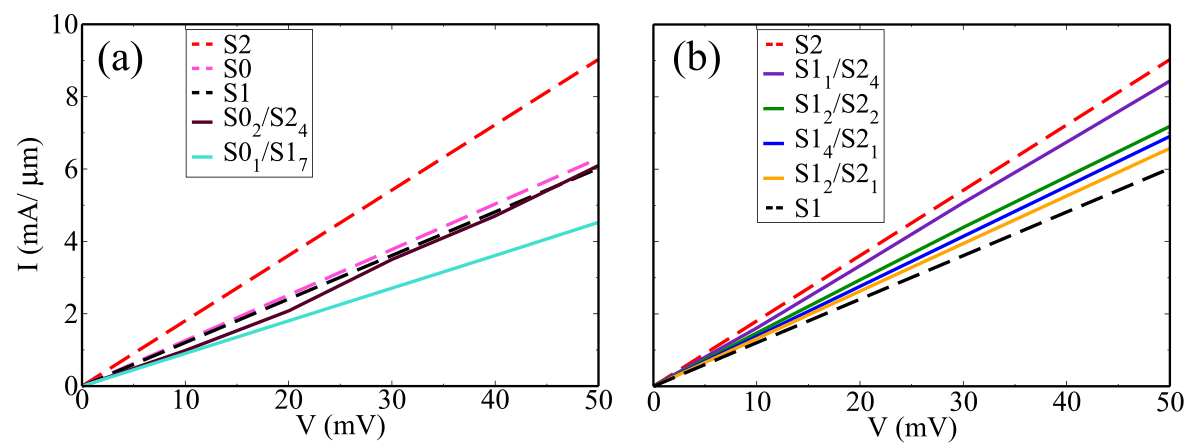}
\caption{\label{correntes2} The calculated electric current as a 
function of the bias voltage applied in the $y$ direction for S0/S1, and 
S0/S2 (a), and S1/S2 (b).}
\end{figure}

In Fig.\,\ref{trans-s0}(a1) and (b1) we present the setup 
used for the electronic transport calculations in S0$_1$/S1$_7$ and 
S0$_2$/S2$_4$. Our results of transmission probability ($T$) along the 
$y$-direction of  S0$_1$/S1$_7$ ($T_y^{S0_1/S1_7}$) and  S0$_2$/S2$_4$ 
($T_y^{S0_2/S2_4}$), Fig.\,\ref{trans-s0}(a2) and (b2), reveal that near the 
Fermi level the transmission probabilities of  the BSLs are lower when 
compared with those  of the pristine phases, {\it viz.}: $T_y^{\rm 
S0_1/S1_7}<T_y^{\rm 
S1}$ and $T_y^{\rm S0_2/S2_4}<T_y^{\rm S2}$. Such a reduction of the $T$ is 
due to the formation  of localized  states, characterized by the 
dispersionless 
energy bands along the YS directions [Fig.\,\ref{bandas-proj-mix}(a3) and (b3)] 
upon the formation of S0/S1 and S0/S2 interfaces. That is, there is an 
interaction between the propagating waves and the 
electronic states with wave vectors along the YS direction.
Meanwhile, as shown Fig.\,\ref{trans-s0}(a2) and (b2), the transmission 
probabilities along the $x$ direction, that is, perpendicularly to the 
electronic stripes of S0$_1$/S1$_7$ and S0$_2$/S2$_4$ are nearly zero.  

Our electronic transport results for the S1/S2 BSLs are 
summarized in Fig.\,\ref{trans-s1s2}. The simulation setups used for the 
calculation of $T_y$ in S1$_4$/S2$_1$  ($T_y^{4,1}$) and  S1$_2$/S2$_2$   
($T_y^{2,2}$) are shown  in Fig.\,\ref{trans-s1s2}(a1) and (c1), respectively. 
We find that $T_y^{4,1}$ lies between the transmission probabilities of  
pristine S1 ($T_y^{\rm S1}$) and S2 ($T_y^{\rm S2}$) phases, namely $T_y^{\rm 
S1} < T_y^{4,1} <  T_y^{\rm S2}$, for an energy interval of $E_{\rm F}\pm 
0.4$\,eV [Fig.\,\ref{trans-s1s2}(a2)]. Based upon the projected energy bands, 
Figs.\,\ref{bandas-proj-s1s2-c}(a2) and (a3), we can infer that the larger 
values  of $T_y^{4,1}$, compared with $T_y^{\rm S1}$, can be attributed to the 
strengthening of the $\pi$-hybridization along the inner sites of S1 and the 
presence metallic bands along S2, note that $T_y^{\rm S2} > 
T_y^{\rm S1}$. On the other hand,  the  reduction of $T_y^{4,1}$ below the 
Fermi level, indicated by an arrow in  Fig.\,\ref{trans-s1s2}(a2), is correlated 
with the formation of localized (B-$2p_z$) occupied states with wave vectors 
along the YS direction [Fig.\,\ref{bandas-proj-s1s2-c}(a2) and (a3)] just below 
the Fermi level.

By reducing the width of S1, S1$_4$/S2$_1$\,$\rightarrow$\,S1$_2$/S2$_1$,  we 
find (i)  a reduction of the transmission probability, namely 
$T_y^{2,1}<T_y^{4,1}$, (ii) being larger than  $T_y^{S1}$ only near the Fermi 
level [(i) and (ii) are indicated in  Fig.\,\ref{trans-s1s2}(b)].  In fact, 
these features are  in agreement with the reduction on the density of  
(B-$2p_z$) metallic bands along S1, as shown in 
Fig.\,\ref{bandas-proj-s1s2-c}(b). Meanwhile, the presence of  localized states 
along the YS direction gives rise to the  transmission valley 
below the Fermi level, indicate as (iii) in Fig.\,\ref{trans-s1s2}(b). 
Increasing the width of the S2 region, 
S1$_2$/S2$_1$\,$\rightarrow$\,S1$_2$/S2$_2$,  the features (i)--(iii) described 
above are  are somewhat preserved. However,  in S1$_2$/S2$_2$  the contribution 
of B-$2p_y$ orbitals, localized on S2, becomes dominant to the formation of 
transport channels [Fig.\,\ref{bandas-proj-s1s2-c}(c2)]. That is,  
$\sigma$-channels start to rule the transmission probabilities. Upon further 
reduction (increase) of S1 (S2),  S1$_1$/S2$_4$, there is an increase on the  
transmission probability near the Fermi level, but now dictated by the $\pi$- 
and $\sigma$-channels along S2, Fig.\,\ref{trans-s1s2}(d), which is in agreement 
with  the  wave function overlap shown in Fig\,\ref{bandas-proj-s1s2-c}(d). 

 Finally, based on the Landauer formula we calculate the 
current density ($I$)  as a function of the bias voltage ($V$) for the pristine 
S0,  S1, and S2 phases ($I^{\rm S0}$, $I^{\rm S1}$, and $I^{\rm S2}$, 
respectively), and  the BSLs. The $I\text{-}V$ curves of those systems are 
characterized by a nearly linear behavior, Fig.\,\ref{correntes2}. Among the 
pristine phases, S2 (S1) presents the higher (lower)  current density. It is 
noticeable that  the  current densities of S0$_1$/S1$_7$ ($I^{\rm S0_1/S1_7}$) 
and  S0$_2$/S2$_4$ ($I^{\rm S0_2/S2_4}$) are lower compared with the ones of  
the pristine phases,  Fig.\,\ref{correntes2}(a); whereas  in S1/S2 BSLs we find 
the  current densities lying between $I^{\rm S2}$ and  $I^{\rm S1}$, 
Fig.\,\ref{correntes2}(b). The latter result can be attributed to the formation 
of neat interface structure  between S1 and S2; while in the former there is a 
strengthening of the scattering processes which can be attributed to the large 
atomic distortions at the S0/S1 and S0/S2 interfaces.

\section{Summary and Conclusions}

Based on {\it first-principles} DFT calculations we have 
studied the electronic and structural properties of  recently synthesized   
pristine borophene sheets with different vacancy concentrations, and borophene 
superlattices (BSLs) composed by combinations of borophene layers with different 
structural phases. Once we confirmed the  metallic character of the  pristine 
systems, through  simulations of X-ray Absorption Near-Edge Structure (XANES) we 
unveil the connection between the electronic properties and the atomic 
arrangement of borophene S0, S1, and S2 phases. We found that each structural 
phase present a particular K-edge X-ray absorption spectrum, and thus, well 
defined  XAS fingerprints. Electronic structure calculations reveal confinement 
effects, giving rise to metallic electronic stripes embedded in BSLs. Further 
electronic transport calculation reveals a strong directional dependence,  where 
we found that the transmission probabilities and the formation  of the transport 
channels are tuned by the m/n periodicity of the BSL. Such a tunning of the  
electronic confinement and transport properties  offer a new and interesting set 
of degree of freedom addressing the design electronic nanoelectronic devices on 
2D platforms.

%
%
%
%

\subsection{Appendix}

\begin{figure}[h]
\begin{center}
\includegraphics[width=\columnwidth]{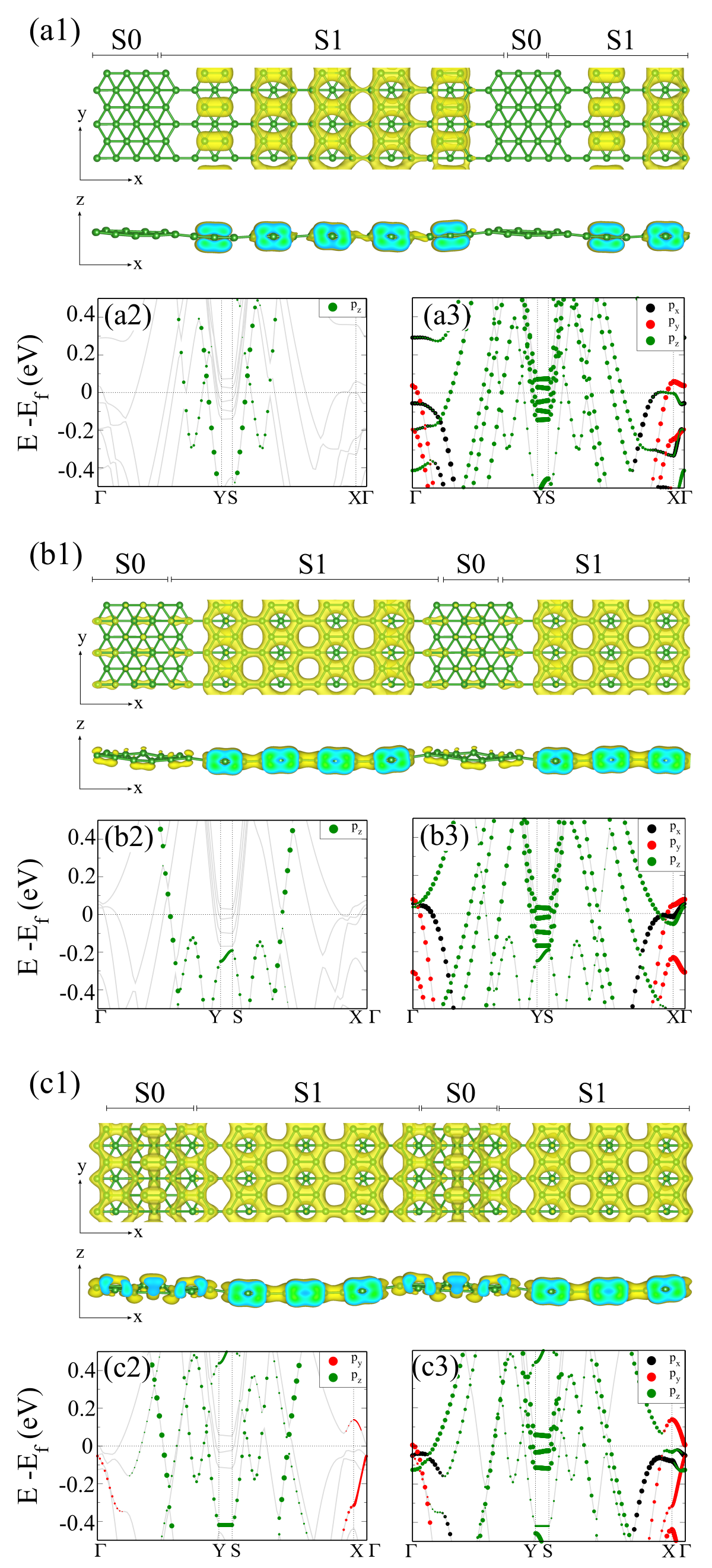}
\caption{\label{band-proj-2-1} Structural models (top-view  and side-view) and 
the projected electronic density of states near the Fermi level, $E_{\rm F}\pm 
0.1$\,eV, of S0$_2$/S1$_6$ (a1), S0$_3$/S1$_5$ (b1), and  S0$_4$/S1$_4$ (c1) 
BSLs. Electronic band structure projeted on the S0 [S1] regions (a2), (b2), and 
(c2) [(a3), (b3), and (c3)]. Isosurfaces of  0.003\,$e$/\AA$^3$ in (a); and 
0.005\,$e$/\AA$^3$ in (b) and (c).}
\end{center}
\end{figure}

\begin{figure}[h]
\begin{center}
\includegraphics[width=\columnwidth]{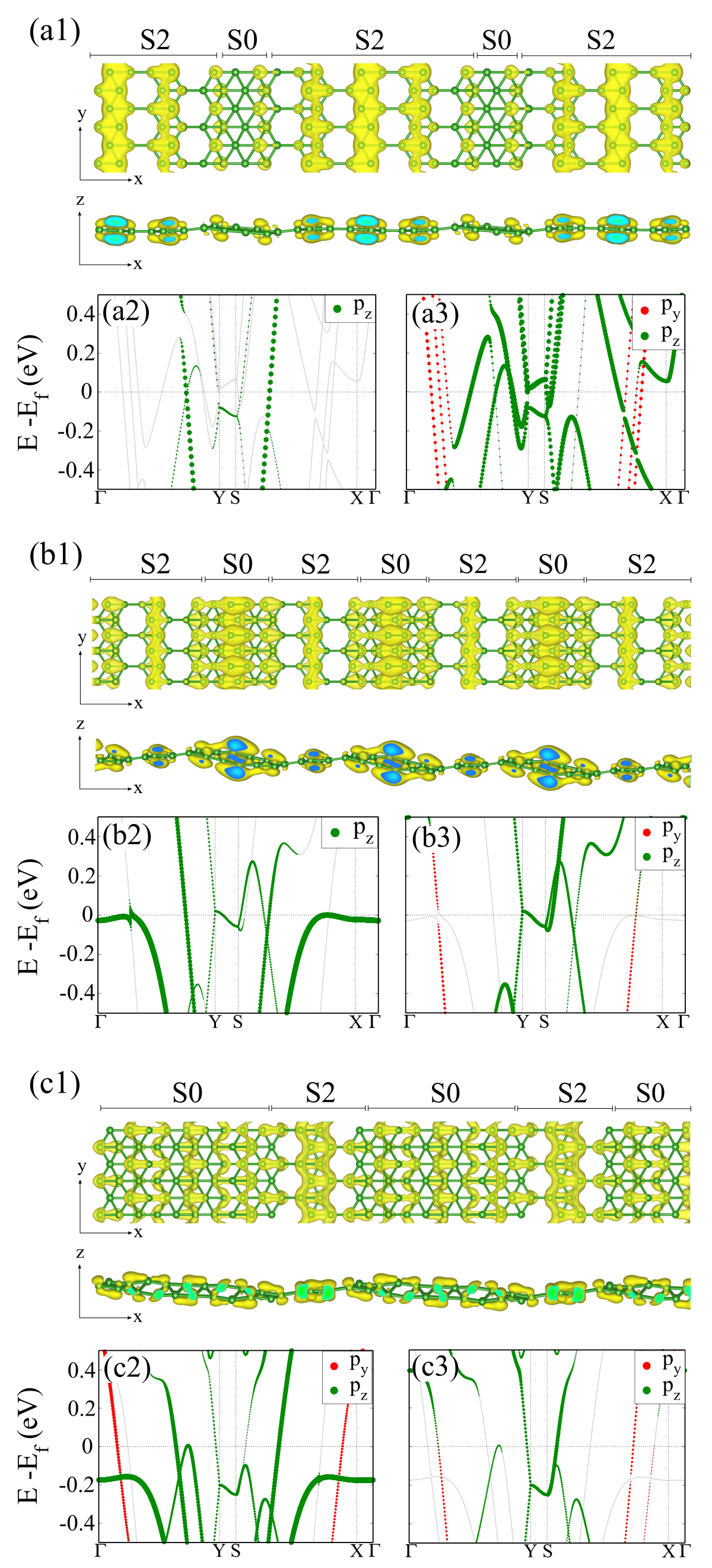}
\caption{\label{band-proj-3} Structural models (top-view  and side-view) and 
the projected electronic density of states near the Fermi level, $E_{\rm F}\pm 
0.1$\,eV, of S0$_2$/S2$_2$ (a1), S0$_4$/S2$_1$ (b1), and  S0$_8$/S2$_1$ (c1) 
BSLs. Electronic band structure projeted on the S0 [S2] regions (a2), (b2), and 
(c2) [(a3), (b3), and (c3)]. Isosurfaces of  0.003\,$e$/\AA$^3$ in (a) and (b);
and 0.0003\,$e$/\AA$^3$ in (c).}
\end{center}
\end{figure}

\begin{figure*}[h]
\includegraphics[width=2\columnwidth]{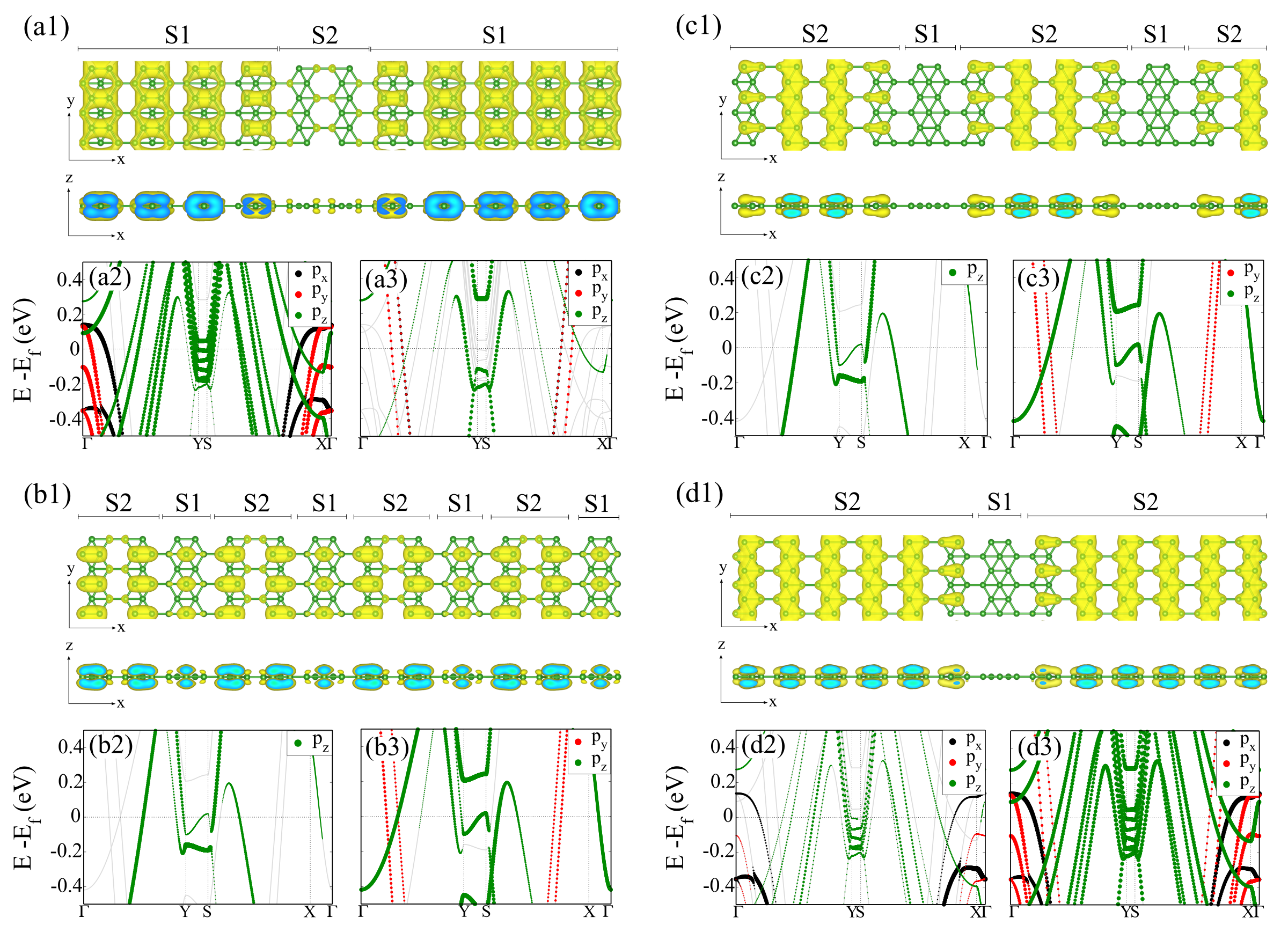}
\caption{\label{banda-proj-4} Structural models (top-view  and side-view) and 
the projected electronic density of states near the Fermi level, $E_{\rm F}\pm 
0.1$\,eV, of S1$_m$/S2$_n$ for m/n = 6/1 (a1), 1/1 (b1), 1/2 (c1), and 1/6 
(d1).
 Electronic band structure projeted on the S1 [S2] regions (a2), (b2),  
(c2), and (d2) [(a3), (b3), (c3), and (d3)]. Isosurfaces of 0.0015\,$e$/\AA$^3$ 
in (a) and (b); and 0.003\,$e$/\AA$^3$ in (c) and (d).}
\end{figure*}

In Figs.\,\ref{band-proj-2-1}(a1)-(c1) we present the 
localization of the electronic states near the Fermi level, $E_{\rm 
F}\pm\,0.1$\,eV, of S0$_{\rm m}$/S1$_{\rm n}$ for m/n = 2/6, 3/5, and 4/4. There 
is an  increase of the vertical buckling along the BSL proportional to the area 
of the S0 region.  However, the localization of the metallic bands along the S1 
region  (mostly due to the B-$2p_z$ orbitals) has been kept 
[Figs.\,\ref{band-proj-2-1}(a2)-(b2) and \ref{band-proj-2-1}(a3)-(b3)], although 
somewhat dimmed by increasing the proportion of S0. Similarly features have been 
found in S0$_2$/S2$_2$, where the surface area of S2$_2$ is larger in comparison 
with that of S0$_2$. On the other hand, the localization of the electronic 
states on S0 increases by increasing it area, Fig.\,\ref{band-proj-3}. In 
Fig.\,\ref{banda-proj-4} we present the electronic band structure, and the 
projection of the electronic states near the Fermi level for the S1$_{\rm 
m}$/S2$_{\rm n}$ BSLs, with m/n =  6/1, 1/1, 1/2, and 1/6, where it is 
noticeable the change on the  localization of the electronic states near the 
Fermi level, namely from S1 in S1$_6$/S2$_1$ [Fig.\,\ref{banda-proj-4}(a)] to 
S2 in S1$_1$/S2$_6$ [Fig.\,\ref{banda-proj-4}(d)].

\section*{Acknowledgements}

The authors acknowledge   financial   support   from   the Brazilian  agencies  
CNPq, and FAPEMIG, FAPES and the  LNCC (SCAFMat2), CENAPAD-SP for 
computer time.

\bibliography{RHMiwa}

\begin{thebibliography}{47}
\bibitem [{Coh()}]{CohesionEnergy}%
  \BibitemOpen
  \href@noop {} {}\bibinfo {note} {$E^c$ is defined as the total energy
  difference between the borophene sheet or superlattice and the isolated B
  atoms.}\BibitemShut {Stop}%
\end{thebibliography}

\end{document}